\documentclass[aps,prb,reprint,superscriptaddress,longbibliography]{revtex4-2}
\usepackage[utf8]{inputenc}
\usepackage{amsmath}
\usepackage{amssymb}
\usepackage{graphicx}
\usepackage[caption=false]{subfig}
\usepackage[colorlinks=true,citecolor=blue,urlcolor=blue]{hyperref}

\newcommand{\avg}[1]{\overline{ #1 }}

\DeclareMathOperator*{\mincut}{mincut}

\newcommand{\papertitle}{Infinite-randomness criticality in monitored quantum dynamics with static disorder}

\begin{document}

\title{\papertitle}

\author{Aidan Zabalo}
\affiliation{Department of Physics and Astronomy, Center for Materials Theory, Rutgers University, Piscataway, New Jersey 08854, USA}

\author{Justin H. Wilson}
\affiliation{Department of Physics and Astronomy, Center for Computation and Technology, Louisiana State University, Baton Rouge, LA 70803, USA}

\author{Michael J. Gullans}
\affiliation{Joint Center for Quantum Information and Computer Science, NIST/University of Maryland, College Park, Maryland 20742, USA}

\author{Romain Vasseur}
\affiliation{Department of Physics, University of Massachusetts, Amherst, Massachusetts 01003, USA}

\author{Sarang Gopalakrishnan}
\affiliation{Department of Physics, The Pennsylvania State University, University Park, Pennsylvania 16802, USA}

\author{David A. Huse}
\affiliation{Department of Physics, Princeton University, Princeton, New Jersey 08544, USA}
\affiliation{Institute for Advanced Study, Princeton, New Jersey 08540, USA}

\author{J. H. Pixley}
\affiliation{Department of Physics and Astronomy, Center for Materials Theory, Rutgers University, Piscataway, New Jersey 08854, USA}

\date{\today}

\begin{abstract}
We consider a model of monitored quantum dynamics with quenched spatial randomness: specifically, random quantum circuits with spatially varying measurement rates. These circuits undergo a measurement-induced phase transition (MIPT) in their entanglement structure, but the nature of the critical point differs drastically from the case with constant measurement rate. In particular, at the critical measurement rate, we find that the entanglement of a subsystem of size $\ell$ scales as $S \sim \sqrt{\ell}$; moreover, the dynamical critical exponent $z = \infty$. The MIPT is flanked by Griffiths phases with continuously varying dynamical exponents. We argue for this infinite-randomness scenario on general grounds and present numerical evidence that it captures some features of the universal critical properties of MIPT using large-scale simulations of Clifford circuits.  
These findings demonstrate that the relevance and irrelevance of perturbations to the MIPT can naturally be interpreted using a powerful heuristic known as the Harris criterion.
\end{abstract}

\maketitle

The study of hybrid dynamics, unitary evolution interspersed with measurements, in random quantum circuits has garnered a significant amount of attention in recent years~\cite{li2019measurement,skinner2019measurement,noel2021observation,PotterVasseurReview}.
It has become well-established that the competition between scrambling dynamics and local measurements leads to a measurement-induced phase transition (MIPT), which is a phase transition in the entanglement structure of the quantum state conditional on a set of (typical) measurement outcomes.
In previously studied models of measurement-induced criticality a recurring theme is the emergence of conformal invariance at the critical point~\cite{li2019measurement,skinner2019measurement,gullans2020dynamical,jian2020measurement,li2021conformal,zabalo2020critical,zabalo2022OperatorSpectrum}.
This symmetry holds even in the presence of fairly drastic modifications of the dynamics, e.g., the addition of conservation laws~\cite{lavasani2021measurement,bao2021symmetry,li2021robust,agrawal2021entanglement,barratt2021field} or the removal of scrambling in lieu of few-site measurements~\cite{ippoliti2021entanglement,lavasani2021topological,van2021entanglement,lang2020entanglement}.
A natural question is whether conformal invariance is robust against the presence of weak spatial randomness---e.g., measurement rates that are spatially varying. 

The MIPT in random circuits has \emph{space-time} randomness, and in the standard treatment~\cite{li2019measurement,skinner2019measurement,gullans2020dynamical,jian2020measurement,li2021conformal,zabalo2020critical,lavasani2021measurement,bao2021symmetry,li2021robust,agrawal2021entanglement,barratt2021field,ippoliti2021entanglement,lavasani2021topological,van2021entanglement,lang2020entanglement,sang2021measurement,choi2020quantum,lunt2021measurement,alberton2021entanglement,szyniszewski2019entanglement,li2018quantum,bao2020theory,lunt2020measurement,goto2020measurement,tang2020measurement,cao2018entanglement,nahum2021measurement,turkeshi2020measurement,zhang2020nonuniversal,szyniszewski2020universality,fuji2020measurement,rossini2020measurement,vijay2020measurement,turkeshi2021measurement,zabalo2022OperatorSpectrum,sierant2022dissipative,sharma2022measurement,chen2021non,han2022measurement,iaconis2020measurement} this randomness is spatially and temporally uncorrelated. As such, the correlation length exponent $\nu$ at this transition is constrained to satisfy the Chayes-Chayes-Fisher-Spencer (CCFS) bound~\cite{chayes1986finite} $\nu \geq 2/D$, where $D = d +1$ is the spacetime dimension. Numerical studies in $d = 1$ have yielded an exponent $\nu \approx 1.3$~\cite{li2019measurement,gullans2020dynamical,zabalo2020critical,agrawal2021entanglement,lavasani2021measurement} that is clearly consistent with the CCFS bound. However, stability against quenched spatial randomness (which is \emph{columnar}, i.e., perfectly correlated in time) imposes the stronger bound $\nu \geq 2/d$, which is not satisfied. 
Consequently, the stability argument due to Harris~\cite{harris1974effect,sachdev2011quantum} implies that the conformally invariant critical point is unstable to any added static spatial randomness. 

In this work we  we explore what new critical universality class this instability leads to.  We pursue two complementary approaches: first, we explore classically simulable Clifford circuits with a spatially varying measurement rate; second, we construct a real-space renormalization-group (RSRG) treatment~\cite{FisherRSRGIsing,FisherIsingPRB,FisherRSRGXXZ,IGLOI2005277,REFAEL2013725} that is strictly valid in the limit of large on-site Hilbert space dimensions, but that may plausibly  describe the fixed point more generally. The two approaches are mutually consistent in many respects, but qualitative deviations between the two critical points are found in some observables. 

The RSRG mapping yields the following key, universal predictions: (i)~the new critical point exhibits activated dynamical scaling, i.e., space and time scale with the relation $\log t \sim \sqrt{x}$ [cf. the relativistic behavior at the conventional MIPT ($x \sim t$)]; (ii)~the steady-state entanglement of a subsystem of size $\ell$ at the critical point scales as $S(\ell) \sim \sqrt{\ell}$ [cf. $\sim \log \ell$ in the absence of static disorder]; (iii)~the MIPT is flanked by Griffiths phases in which the late-time dynamics is governed by rare-region effects.  We put forward these three behaviors as sufficient criteria to claim two critical points are described by the same infinite-randomness fixed point.
Our numerical evidence indicates that both Clifford circuits and percolation are consistent with each of these predictions. However, we also find that the tripartite information at the Clifford critical point acquires a broad distribution, with its average growing as a fractional power law of system size. This numerical observation appears robust, and goes beyond any existing theoretical calculation. This feature also appears to be absent in our numerical simulations of percolation with columnar disorder. 

\begin{figure*}
	\centering
    \subfloat[\label{fig:ent_profile_griff}]{\includegraphics[width=0.4\linewidth]{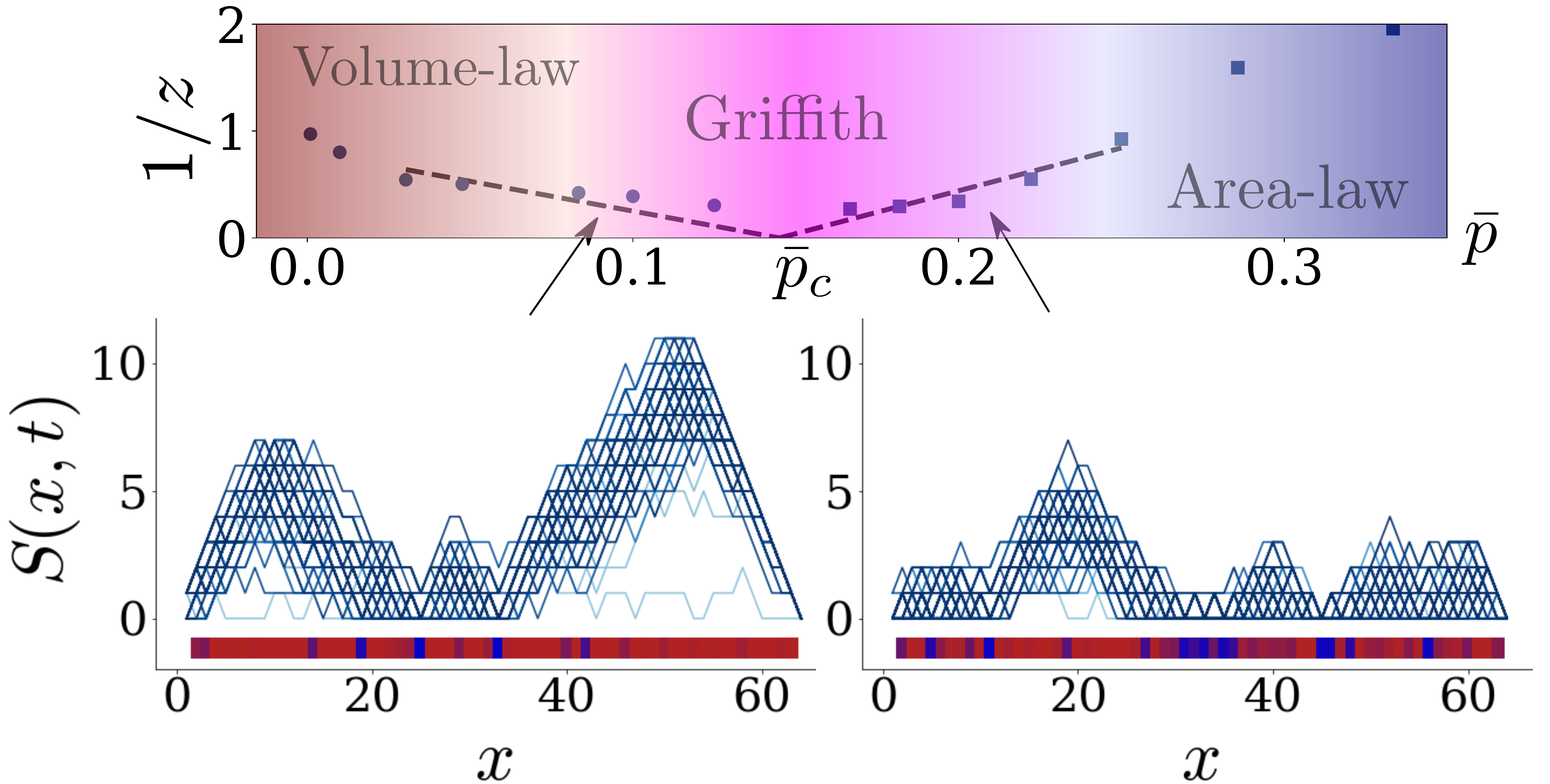}}
    \subfloat[\label{fig:S_z_vol}]{\includegraphics[width=0.3\linewidth]{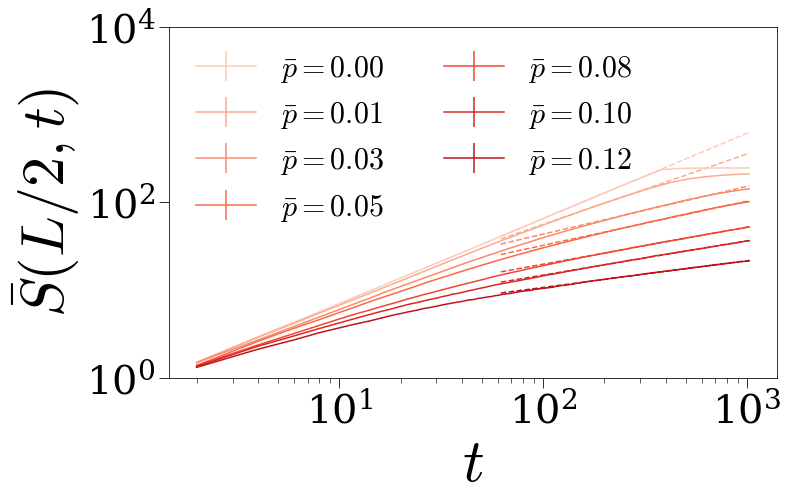}}
    \subfloat[\label{fig:S_z_area}]{\includegraphics[width=0.3\linewidth]{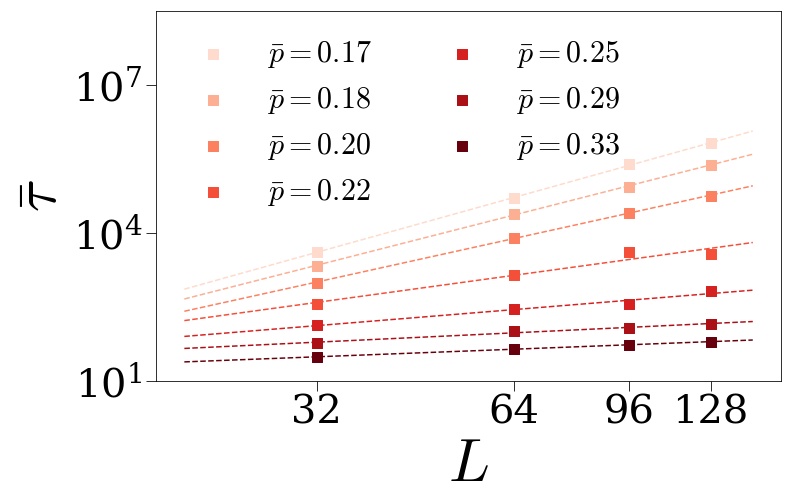}}
    \caption{
        (a) (top) Phase diagram of the disordered circuit model indicating the behavior of the dynamic exponent $z$ near the critical point and Griffith phase. The dynamic exponent $z = z(\avg{p})$ is a continuous function of the average measurement rate $\avg{p}$ and increases as $\avg{p}\to \avg{p}_c$ from the volume-law and area-law phases (data points are obtained from numerical simulations, dashed lines are the result of the RSRG prediction $1/z\sim|p-p_c|^{\nu\psi}=|p-p_c|$).
        (bottom) Entanglement profile $S(x,t)$ as a function of the cut $x$ and time $t$
        (darker blue indicates later times). Each panel corresponds to a single
        trajectory and a different value of $\avg{p}$ in the Griffith regime (left: $\avg{p} = 0.91$, right: $\avg{p} = 0.20$). The red to blue color at the bottom indicates small to large values of of the local measurement probability $p_x$, respectively.
        (b) Early time behavior of the average half cut entanglement entropy, $\avg{S}(L/2,t)$,
        in the volume law phase of the disordered circuit. The data is fit to a
        power law given by $\avg{S}(L/2,t) \sim t^{1/z}$.
        (c) Purification time in the area law phase used to extract the
        dynamical exponent from a fit of $\bar{\tau} \sim L^z$.
    }
\end{figure*}

\emph{Models and entanglement probes of the MIPT}:
We consider random circuit models of the ``brick wall'' structure that consist of randomly chosen two qubit gates between nearest neighbor $q$-state spins. For our numerical simulations we focus on randomly sampling Clifford gates between qubits ($q=2$) that allow us to reach large system sizes as they can be simulated with classical efficiency~\cite{gottesman1998heisenberg,aaronson2004improved,audenaert2005entanglement,fattal2004entanglement}. For our analytic treatment of the problem we take Haar random gates and send $q\rightarrow \infty$. Last, we also consider a classical limit of this model by solving a percolation problem with columnar disorder, in the supplement~\cite{supp}.

To build static spatial randomness into the problem we focus on a  position-dependent measurement probability $p_x$ that is applied between every layer of unitary gates and is given by
$p_x = \mathbf{r}_{x}^n,$
where $\mathbf{r}_{x}$ is a random variable uniformly distributed in $[0, 1]$ and $n$ is the tuning parameter. For a given realization of the circuit we generate a fixed static $p_x$ and find it convenient to parameterize the strength of the measurement rate by its disorder averaged value, i.e. $\avg{p} = \frac{1}{n + 1}$, and in the following we denote averages over circuit realizations via $\avg{\cdots}$.
This distribution is chosen in order to have long tails so that rare regions of atypical values of $p_x$ play an important role in the dynamics for relatively modest system sizes, but we do not expect it to change the universal properties of the critical point in the thermodynamic limit. 
This static measurement profile produces a ``columnar'' disorder pattern in space-time.

The measurement transition is present in quantities that involve averages that are non-linear in the reduced density matrix conditional on measurement outcomes, therefore we focus on entanglement probes of the system. Firstly, we focus on the von-Neumann entanglement entropy by dividing our system (with periodic boundary conditions) into two regions $A$ (of length $x$) and $B$ (of length $L-x$) and trace out region $B$, resulting in the entanglement entropy $S(x,t)=-\mathrm{Tr}_A[\rho_A\log_2\rho_A]$, where $\rho_A=\mathrm{Tr}_B|\psi(t)\rangle\langle \psi(t)|$ is a trace of the density matrix over region $B$ where $|\psi(t)\rangle$ is the time-evolved wavefunction on a given sampled quantum trajectory.

Second, we utilize the behavior of the tripartite mutual information to identify and characterize the critical point of the entanglement transition.
The tripartite mutual information is defined as
\begin{equation}
  \begin{split}
    \mathcal{I}_{3}(A,B,C) \equiv S(A) + S(B) + S(C) - S(A\cup B)&\\
    - S(A\cup C) - S(B\cup C) + S(A\cup B\cup C)&
  \end{split}
\end{equation}
where we have omitted the time label and have partitioned the geometry of the system into
 adjacent regions $A,B,C$ each of size $L/4$. The value of $\mathcal{I}_3$ depends sensitively on the nature of the circuit realization; we thus have a probability distribution $P[\mathcal{I}_3]$ over trajectories and circuit realizations, and we denote the circuit-averaged value as $\avg{\mathcal{I}_3}$. 
In the presence of strong spatial randomness the mean  $\avg{\mathcal{I}_3}$ is not  representative of the distribution $P[\mathcal{I}_3]$ as it becomes broad with fat tails requiring  a more general scaling ansatz  to capture the critical dynamics. Thus we study the distribution $P[\mathcal{I}_3]$ in detail, which motivates us to generalize the scaling ansatz from Refs.~\cite{zabalo2020critical,gullans2020dynamical} to include the possibility of extensive scaling at the critical point (see supporting distribution data in the Supplement~\cite{supp}).
As a result, in order to numerically identify the critical point we find it most accurate  to work with the distribution $P[\mathcal{I}_{3}]$.

To understand the nature of purification dynamics we also study the ancilla order parameter, which is defined as follows~\cite{gullans2020scalable}:
At $t=0$, a site in the system is maximally entangled with a reference qubit and the system is scrambled by unitary evolution for a time $2L$.
The system is then evolved under the hybrid dynamics for an additional time $2L$ and the average entanglement entropy of the reference qubit, $\avg{S}_Q$ acts as an order parameter for the transition.
In the volume-law phase, local measurements do not reveal information about the reference qubit and the entanglement entropy remains nonzero up to times exponential in the system size while in the area-law phase, the measurements quickly collapse the state of the reference qubit and disentangle it from the system.
Near the critical point, $\avg{S}_Q$ obeys single parameter scaling allowing for an additional probe of the transition that is complimentary to $\mathcal{I}_3$.

\emph{RSRG for quenched disorder}: The transition is analytically tractable for Haar-random circuits in the limit of large on-site Hilbert space $q\to \infty$, using mappings onto replica statistical mechanics models~\cite{RTN2019,AdamTianci,bao2020theory,jian2020measurement,nahum2021measurement,CliffordStatMech}. Upon averaging over random Haar gates, measurement locations and outcomes, any nonlinear function of the density matrix of the system can be mapped onto an effective two-dimensional $k!$-state Potts model in the replica limit $k \to 1$, which is known to describe {\em bond percolation}~\cite{bao2020theory,jian2020measurement}. Quenched disorder in the circuit leads to {\em columnar disorder} in the statistical mechanics model, which is amenable to RSRG techniques~\cite{FisherRSRGIsing,FisherRSRGXXZ,supp}. We find that for any number of replicas, and directly in the replica limit $k \to 1$, the transition is described by an infinite randomness fixed point with space-time scaling $\log t \sim \ell^{\psi}$, with $\psi=\frac{1}{2}$ ($z= \infty$ that diverges like $z\sim \ell^{\psi}/\log\ell$), in agreement with known results on percolation with columnar disorder~\cite{PercolationRandom}. Scaling properties follow from known results~\cite{FisherRSRGIsing,FisherIsingPRB,FisherRSRGXXZ}: in particular the (average) correlation length exponent is $\nu=2$, and the scaling in the phases is controlled by Griffiths effects.

One important consequence of the mapping is the prediction that the steady-state entanglement entropy at criticality scales as $\avg{S} \sim \sqrt{\ell}$. In the statistical mechanics picture, the entanglement entropy is related to the free energy cost of inserting a domain wall of size $\ell$ at the (spatial) boundary of the system~\cite{bao2020theory,jian2020measurement}. The free-energy cost of a domain wall is related to the \emph{logarithm} of the boundary two-point function, which typically scales as $\exp(-\ell^{\psi})$~\cite{FisherRSRGIsing,FisherIsingPRB,FisherRSRGXXZ}. Since $\avg{S}$ is related to the logarithm of the boundary two-point function it is dominated by typical samples and not rare samples; hence the result above. From the spacetime scaling mentioned above it follows that $\avg{S} \sim \log t$. 

Although our predictions are restricted to $q=\infty$, infinite randomness fixed points tend to be ``superuniversal''~\cite{SenthilMajundarPotts,DamleHuse}: for example, the critical properties of the random Potts model do not depend on the number of states~\cite{SenthilMajundarPotts}, in sharp contrast with the clean case. It is therefore plausible to expect those specific predictions to extend to the finite $q$ case as well, as we verify numerically below for qubits ($q=2$).  Notably, however, we find deviations between the universal behavior of the $q = \infty$ percolation model and the Clifford model in the scaling of  mutual information quantities at the critical point.

\begin{figure}
    \centering
    \includegraphics[width=0.95\linewidth]{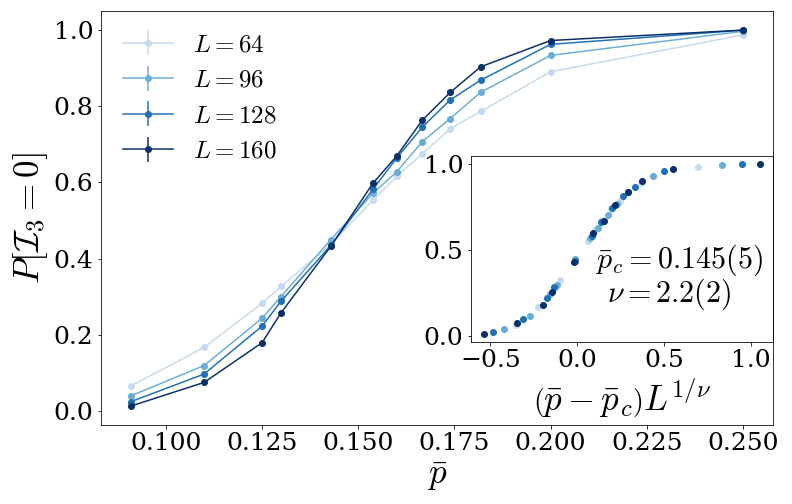}
    \caption{\label{fig:pI3eq0}
        Data and collapse of the distribution $P[\mathcal{I}_3 = 0]$ used to determine critical point $\avg{p}_c$ and the correlation length exponent $\nu$.
    }
\end{figure}

\begin{figure*}
    \centering
    \subfloat[\label{fig:SvSqrtL}]{\includegraphics[width=0.32\linewidth]{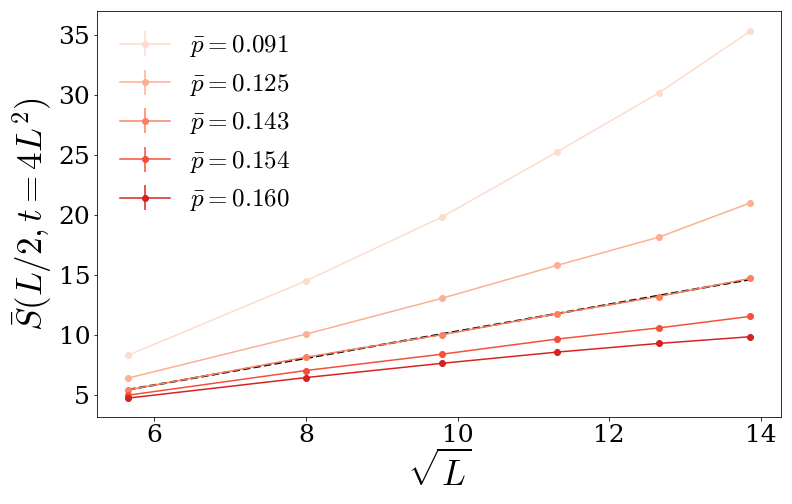}}
    \subfloat[\label{fig:tstar}]{\includegraphics[width=0.32\linewidth]{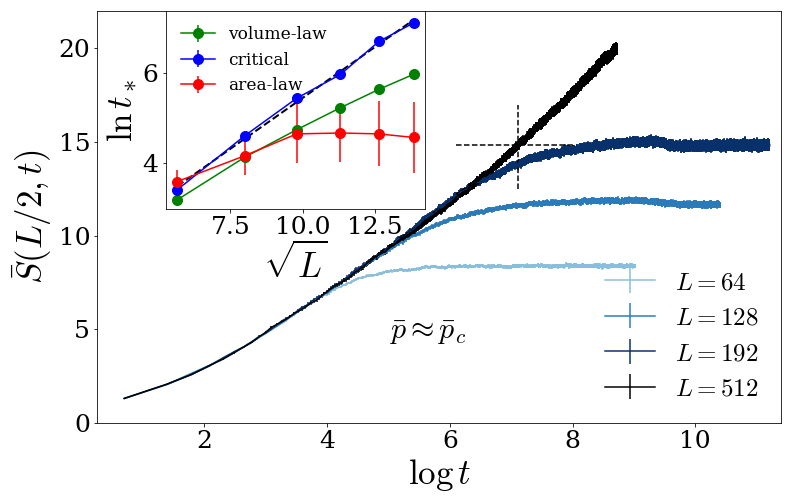}}
    \subfloat[\label{fig:opvt}]{\includegraphics[width=0.32\linewidth]{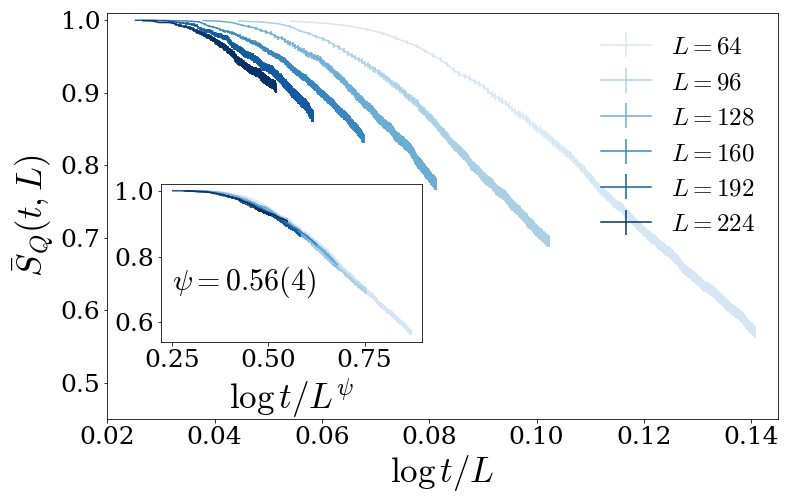}}
    \caption{
        (a) Average half-cut entanglement entropy at late times for various measurement rates.
        Near the critical point, the entanglement entropy behaves as $\avg{S}\sim\sqrt{L}$ as shown by the black dotted line fit for $\avg{p}_c \approx 0.14$.
        (b) Example fit (at $\avg{p} \approx \avg{p}_{c}$) used to extract the saturation time $t_*$ as
        a function of system size.
        (inset) In the volume law phase, we find a stretched exponential behavior
        $t_* \sim \exp[\sqrt{L}]$ while in the area law phase it approaches a constant.
        (c) Order parameter dynamics for the disordered measurement rate. The data
        collapses onto a single universal curve using the activated dynamic
        scaling ansatz.
    }
\end{figure*}

\emph{Griffiths phases}:
The RSRG treatment predicts that on either side of the critical point, certain dynamical quantities are dominated by rare-region effects. 
The presence of rare region effects is manifest in spatial profiles of the entanglement entropy $S(x,t)$ for cuts at various positions $x$ and times $t$ for a given profile of measurement rates, as depicted in Fig.~\ref{fig:ent_profile_griff}. 
At small values of $\avg{p}$ we find that frequently measured regions act as
bottlenecks that hinder the growth of the entanglement past that cut. In contrast, for large values of $\avg{p}$ we see that regions that are measured infrequently produce highly scrambled local regions.

The observables that quantitatively diagnose these Griffiths effects are distinct in the two phases. In the volume-law phase, we expect regions with a high measurement probability to act as bottlenecks for entanglement growth. 
Consider a region of size $\ell$ that is locally in the area-law phase. Suppose the region gets entangled with degrees of freedom to its left, so it is in a mixed state. Measurements rapidly purify this mixed state; the probability that it remains mixed long enough for entanglement to spread across it is suppressed as $e^{-\ell/\xi}$ where $\xi$ is the local correlation length inside the rare region. Therefore the rate at which entanglement spreads across a rare region of size $\ell$ scales as $e^{-\ell/\xi}$. Because the measurement rate is spatially uncorrelated, the density of rare regions of size $\ell$ is exponentially suppressed, as $f^\ell$ for some $f$ that depends on the microscopic details of the disorder but approaches unity at the transition. Therefore, bottlenecks that allow entanglement growth at rate $\leq \gamma$ occur at density $f^{-\xi \log \gamma} \sim \gamma^{\alpha}$, where $\alpha\equiv \xi |\log f|$. Ref.~\cite{nahum2018dynamics} addressed the problem of entanglement growth in the presence of a power-law distribution of bottlenecks; it was found that $\avg{S(t)} \sim t^{1/z}$, where $z = (\alpha+1)/\alpha$.

In Fig.~\ref{fig:S_z_vol} the late time entanglement growth is shown along with 
a power-law fit to at late times to extract $z$ in the volume law phase. We stress that this in stark contrast to space-time random circuits that scale ballistically in time.

In the area-law phase, rare locally scrambling regions do not dominate the steady-state entanglement; the observable they dominate, instead, is the purification rate of an initially mixed state. A region of size $\ell$ that is locally in the volume law phase purifies on a timescale $\sim e^{\ell/\xi}$ where $\xi$ is the local correlation length of the region. As before, the density of volume law regions of size $L$ is suppressed as $\tilde{f}^\ell$ for some $\tilde{f}$ that approaches unity at the transition. In a sample of size $L$, the largest expected volume-law region has $\tilde{f}^\ell \sim 1/L$ so $\ell \sim \log L / |\log \tilde{f}|$. The purification time of the sample is controlled by this largest bottleneck and therefore scales as $\tau(L) \sim L^z$ where $z = {1/(\xi |\log \tilde f|)}$. 
Deep in the area law phase $z\to 0$ but as $\avg{p}\to \avg{p}_{c}$ rare region effects begin to play a role in the dynamics and the size of the largest rare region determines the purification time, giving rise to the power-law behavior $\avg{\tau}\sim L^z$. In Fig.~\ref{fig:S_z_area},  $z$ is extracted via fits to the largest system sizes.

In summary, we find that in each Griffith regime the dynamic exponent $z=z(\avg{p})$ is a continuously varying function of $\avg{p}$. From the volume-law side, $z$ starts near 1 and increases as $\avg{p}\to \avg{p}_c$ while from the are law side $z$ starts near 0 and increases, see Fig.~\ref{fig:ent_profile_griff}.
These results provide an underestimate of $z$ near $\avg{p}_c$ as it is heavily affected by finite-size effects.

\emph{Identifying the critical point and its properties}:
Next, we turn to determining the location of the critical point using the tripartite mutual information. 
As previously mentioned, due to the distribution $P[\mathcal{I}_3]$ developing fat tails, see supplement~\cite{supp}, the mean $\avg{\mathcal{I}}_3$ does not fully characterize the distribution, which dramatically modifies the single parameter scaling $\avg{\mathcal{I}}_3$ near the transition. Therefore, we turn to properties of
$P[\mathcal{I}_3]$ to identify $\avg{p}_c$. In the volume law phase, the probability to find $\mathcal{I}_3=0$ must vanish in the thermodynamic limit, whereas it must approach unity deep in the area law phase~\cite{supp}.
As shown in Fig.~\ref{fig:pI3eq0}, this behavior is consistent with our numerical data allowing us to identify  $P[\mathcal{I}_3=0]$ as a scaling variable, which importantly  does not require knowing the scaling of $\avg{\mathcal{I}}_3$
only the assumption that it crosses at $\avg{p}_c$. Using the scaling ansatz
\begin{equation}
	P[\mathcal{I}_3=0] \sim F[L^{1/\nu}(\avg{p}-\avg{p}_c)]
\end{equation}
where $F(x)$ is arbitrary scaling function,
we find excellent data collapse as shown in Fig.~\ref{fig:pI3eq0}(inset) for $\avg{p}_c \approx 0.145(5)$ and
$\nu \approx 2.2(2)$ in good agreement with the RSRG. Importantly, the estimated value of $\nu$ is stable with respect to the Harris/CCFS bound.

We would also like to comment on identifying the transition using a data collapse of $\avg{\mathcal{I}}_3$ with the less constrained scaling ansatz $\avg{\mathcal{I}}_{3} \sim L^{a}g\left[(p-p_{c})L^{1/\nu}\right]$, to account for any possible $L$ dependence at the critical point, that could be incurred for example due to a fat tail in $P[\mathcal{I}_3]$. For completeness,
we consider three cases of the scaling function motivated by generality, the behavior of $P[\mathcal{I}_{3}=0]$, and the RSRG picture, see supplementary information for details~\cite{supp}.
In all cases we find similar results for $\avg{p}_{c}$, $\nu$, and $a>0$ with differences $\lesssim 10\%$. Last, we note that the non-zero value of $a$ is beyond the classical limit of the model as we show in the supplement~\cite{supp}.

Motivated by these results, we study the average, half-cut, bipartite entanglement entropy
$\avg{S}(L/2,t)$  as a function of the system
size $L$ and time $t$. Near the transition, for $\avg{p}\approx \avg{p}_c$, in the long time limit ($t\gg L$) we find 
\begin{equation}
    \avg{S}(L/2,t\to\infty)\sim\sqrt{L}
    \label{eqn:SL}
\end{equation}
whereas, for large system sizes ($L\gg t$) we obtain
\begin{equation}
    \avg{S}(L/2\to\infty,t)\sim \log{t},
     \label{eqn:St}
\end{equation}
 as shown in Fig.~\ref{fig:SvSqrtL}
in excellent agreement with the RSRG predictions. These results  demonstrate that the critical point has a divergent dynamic exponent consistent with an infinite randomness fixed point.
In the classical limit of percolation, we have also found the scaling in Eqs.~\eqref{eqn:SL} and \eqref{eqn:St}, see supplement~\cite{supp}.
Additionally, we  compute the saturation time $t_*$ at which $\avg{S}(L/2,t)$
reaches its late time value as shown in Fig.~\ref{fig:tstar}.
In the disordered system,  rare regions will cause the entanglement to grow
sub-ballistically so that the saturation time is no longer $t_* \sim L$.
Numerically, in the volume-law phase we find a stretched exponential $t_* \sim
e^{\sqrt{L}}$ while in the area law phase it approaches a constant, see
Fig.~\ref{fig:tstar} (inset).

Finally, we examine the average order parameter dynamics $\avg{S}_{Q}(t,L)$  at the critical point as shown in Fig~\ref{fig:opvt}. Our results have demonstrated this critical point is of the infinite randomness type that has a divergent dynamic exponent $z\sim\xi^\psi\sim|\avg{p}-\avg{p}_c|^{-\nu \psi}$, therefore we use the activated dynamic scaling ansatz~\cite{fisher1987activated} that yields
\begin{equation}
  \avg{S}_{Q}(t,L) \sim g\left(\frac{\log t}{L^{\psi}}\right),
  \label{eq:sq_ansatz2}
\end{equation}
where $g(x)$ is an arbitrary scaling function and
 $\psi$ is the so-called activation or barrier exponent.
We find the excellent data collapse that yields
$\psi = 0.56(4)$
in reasonable agreement with the RSRG result,
see Fig.~\ref{fig:opvt} (inset).
Importantly, this value of $\psi$ is consistent with the length-time scaling of the entanglement entropy in Eqs.~\eqref{eqn:SL} and \eqref{eqn:St}.

\emph{Discussion}:
Introducing static disorder to the measurement induced phase transition is a relevant perturbation that produces a flow to an (apparently) infinite-randomness critical point. We have constructed a field theoretic description of this transition in terms of a real space renormalization group approach and verified its key predictions using large scale simulations of Clifford circuits as well as its classical limit through percolation.
Our results for the tripartite mutual information for Clifford circuits and simulations of percolation
 show qualitatively different behavior, raising the possibility that these two infinite-randomness fixed points belong to distinct universality classes. Analytically computing this quantity within the RSRG is an important task for future work.

Finally, the results for static randomness presented above stand in stark contrast with the case of static quasiperiodic modulation in space.
The bound governing the relevance of quasiperiodic perturbations added to random circuits is the weaker Luck~\cite{luck1993classification} bound $\nu \geq 1/d$. Therefore quasiperiodic spatial modulations of the measurement rate leave the universal nature of the MIPT unchanged, as we demonstrate in the supplement~\cite{supp}.
The demonstration of a successful application of the Harris/Luck criteria to measurement induced criticality provides a powerful heuristic to interpret relevant and irrelevant perturbations on this information-theoretic transition.
Under this paradigm, future work could begin to understand how topologically-ordered phases and transitions change with static randomness~\cite{lavasani2021measurement}.

\emph{Acknowledgements}:
AZ was supported by a HEERF graduate fellowship, JHP and RV were supported by the Alfred P. Sloan Foundation through Sloan Research Fellowships. We acknowledge support from NSF Grants No. DMR-2103938 (SG), DMR-2104141 (RV), QLCI grant OMA-2120757 (MJG, DAH). RV thanks A.C. Potter and S.A. Parameswaran for discussions of RSRG. 
The authors acknowledge the Beowulf cluster at the Department of Physics and Astronomy of Rutgers University and the Office of Advanced Research Computing (OARC) at Rutgers, The State University of New Jersey (http://oarc.rutgers.edu) for providing access to the Amarel cluster, and associated research computing resources that have contributed to the results reported here.
Part of this research was done using services provided by the OSG Consortium \cite{osg07,osg09}, which is supported by the National Science Foundation awards \#2030508 and \#1836650.

\bibliography{references}

\pagebreak
\widetext
\clearpage

\textbf{\large Supplemental Material: \papertitle}

\setcounter{equation}{0}
\setcounter{figure}{0}
\setcounter{table}{0}
\setcounter{page}{1}
\renewcommand{\theequation}{S\arabic{equation}}
\setcounter{figure}{0}
\renewcommand{\thefigure}{S\arabic{figure}}
\renewcommand{\thepage}{S\arabic{page}}
\renewcommand{\thesection}{S\arabic{section}}
\renewcommand{\thetable}{S\arabic{table}}
\makeatletter

\section{Quasiperiodic measurement profile}
In systems where aperiodic structures are introduced, the Luck criterion~\cite{luck1993classification} states that the aperiodicity is relevant when
\begin{equation}
  \omega > \omega_{c} = 1 - \frac{1}{\nu d},
\end{equation}
where $0 \le \omega < 1$ is the wandering exponent.
For quasiperiodic structures $\omega = 0$ and quasiperiodicity is irrelevant when $\nu \ge 1/d$.
This condition is satisfied for our model and we should, therefore, not expect the universality class of the model to change upon introducing the quasiperiodic measurement profile.

In the quasiperiodic model, the measurement probability on site $x$ given by
\begin{equation}
p_x = \biggr\lvert p\cos\left(\frac{F_{n-2}}{F_{n}}2\pi x\right)\biggr\rvert,
\end{equation}
where $p$ is the tuning parameter and $F_{n}$ and $F_{n-2}$ are Fibonacci numbers from the sequence
\begin{equation}
    F_{n} =  4, 4, 8, 12, 20, 32, 52, 84, 136, 220, 356, 576, ...
  \end{equation}
Using $F_{n} = F_{n-1} + F_{n-2}$ and $\lim_{n\to\infty}\frac{F_{n+1}}{F_{n}} = \varphi$, where $\varphi$ is the irrational number known as the golden ratio, the period $T_{n}= \frac{F_{n}}{F_{n-2}}$ approaches the irrational number $1 + \varphi$.

\begin{figure*}[b!]
	\centering
    \subfloat[\label{fig:cosfib_I3}]{\includegraphics[width=0.3\linewidth]{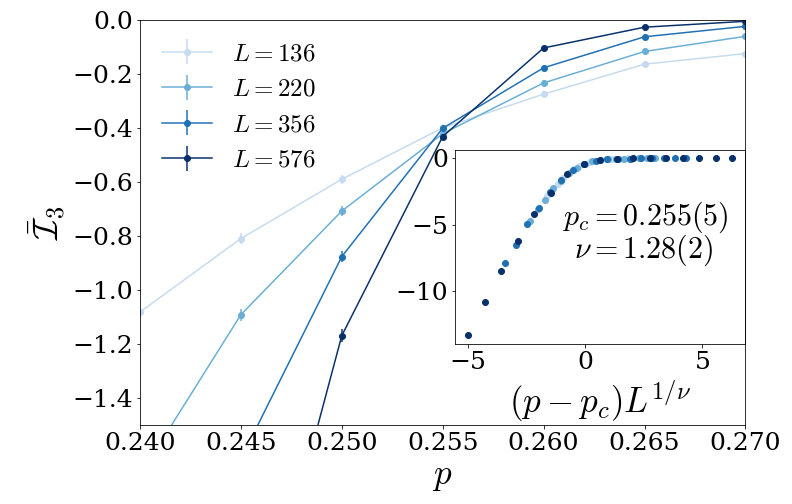}}
    \subfloat[\label{fig:cosfib_op}]{\includegraphics[width=0.3\linewidth]{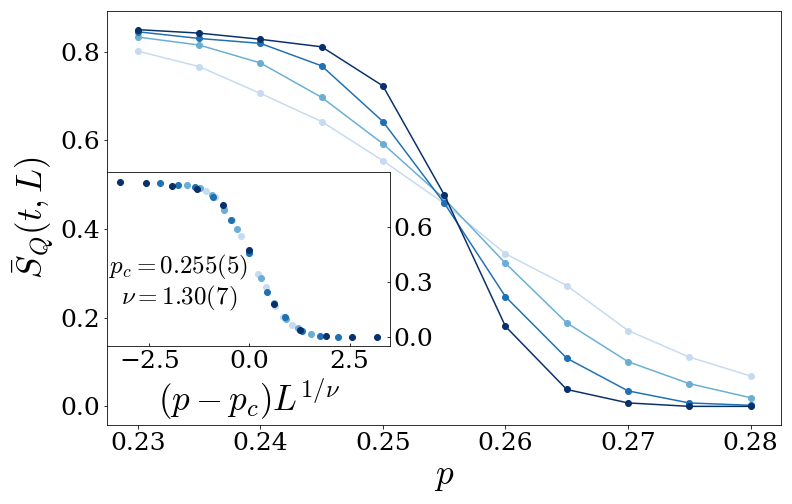}}
    \subfloat[\label{fig:cosfib_z}]{\includegraphics[width=0.3\linewidth]{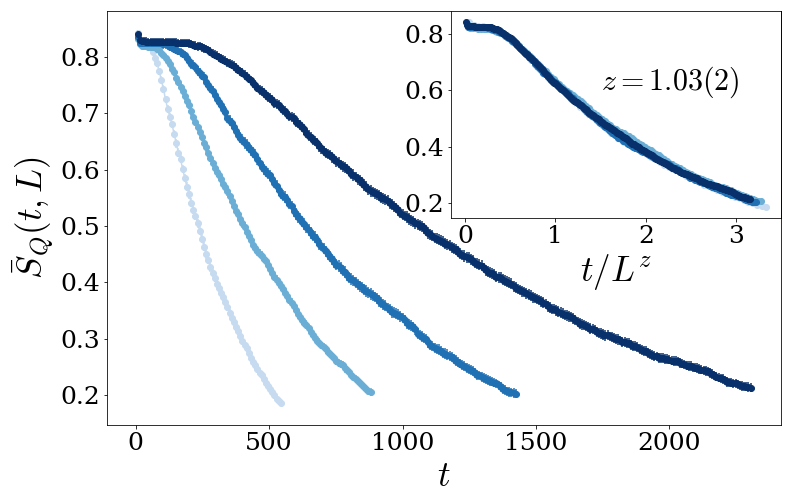}}
    \caption{
    (a) Data collapse of the tripartite mutual information $\avg{\mathcal{I}}_3$ for the quasiperiodic circuit.
    The critical point $p_c$ is shifted but the value of $\nu$ remains unchanged as expected from the Luck criterion.
    (b) Order parameter $\avg{S}_Q(t,L)$ and data collapse used to determine $p_c$ and $\nu$.
    The estimated values for $p_c$ and $\nu$ match the result from $\avg{\mathcal{I}}_3$.
    (c) Dynamics of the order parameter at $p=p_c$ and data collapse used to determine the dynamical exponent $z$ of the quasiperiodic circuit.
    }
\end{figure*}

We begin by identifying the critical point of the entanglement transition through finite-size scaling of the tripartite mutual information using the scaling ansatz
\begin{equation}
    \avg{\mathcal{I}}_3 \sim h[L^{1/\nu}(p - p_c)].
\end{equation}
We find that for $p_c = 0.255(5)$ and $\nu = 1.28(2)$ the data collapses onto a single curve, see Fig.~\ref{fig:cosfib_I3}.
Compared to the traditional model where $p_{x} = p = \mathrm{const.}$, the value of $p_c$ has increased as expected due to the average measurement rate having been reduced by the cosine modulation, i.e., at each site $p_x \le p$ so that $\avg{p}_{x}\to 2p/\pi$.
On the other hand, the value of $\nu$ matches well with previous results suggesting the universality class is unchanged~\cite{li2019measurement,gullans2020dynamical}.

A similar result is found using the entanglement entropy of a reference qubit, $\avg{S}_Q$, as an order parameter~\cite{gullans2020scalable}.
Applying a scaling collapse of the data with ansatz
\begin{equation}
    \avg{S}_Q(t,L) \sim g[L^{1/\nu}(p-p_c), t/L^z]
\end{equation}
we find $p_c = 0.255(5)$ and $\nu = 1.30(7)$, see Fig.~\ref{fig:cosfib_op}.
Additionally, the order parameter dynamics at the critical point, $\avg{S}_Q(t)$, can be used to estimate the dynamical exponent $z$~\cite{gullans2020scalable}.
Fig.~\ref{fig:cosfib_z} shows the data collapses onto a single curve for $z = 1.03(2)$, which is consistent with conformal invariance at the critical point.
These results are consistent with the expectation from the Luck criterion that the universality class is unaffected by the introduction quasiperiodic structure.

\section{Disordered measurement profile}
\subsection{Distribution of \texorpdfstring{$\mathcal{I}_3$}{I3} in the disordered model}
\begin{figure}[b!]
	\centering
    \subfloat[]{\includegraphics[width=0.3\linewidth]{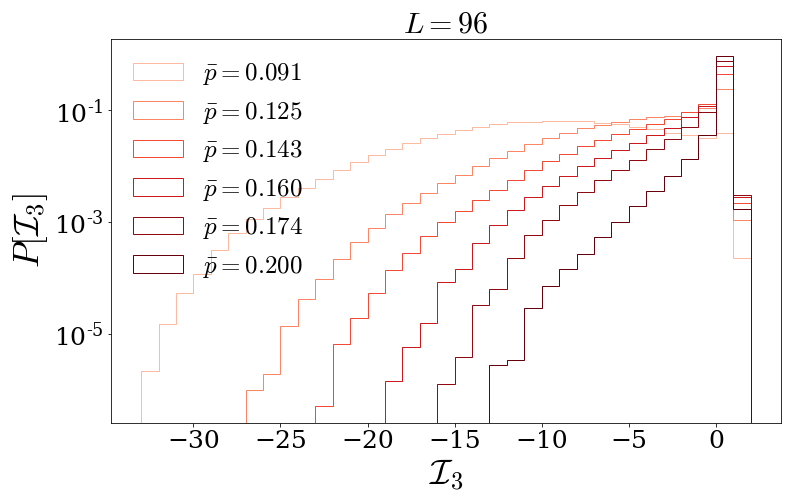}}
    \subfloat[]{\includegraphics[width=0.3\linewidth]{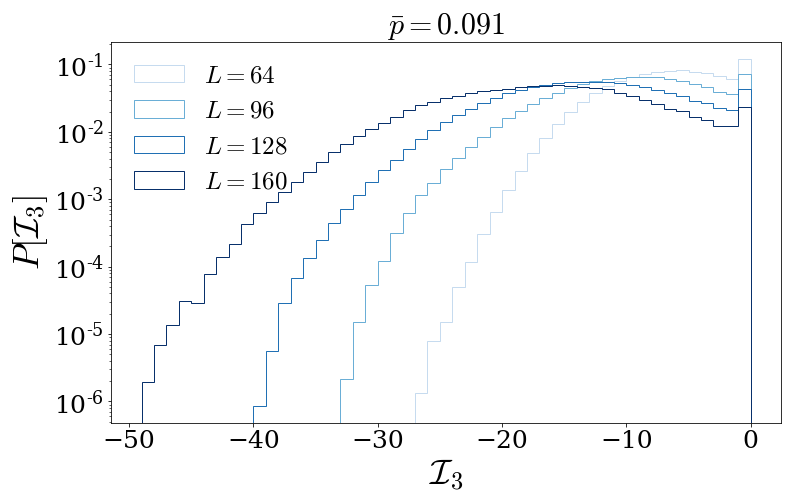}}
    \subfloat[]{\includegraphics[width=0.3\linewidth]{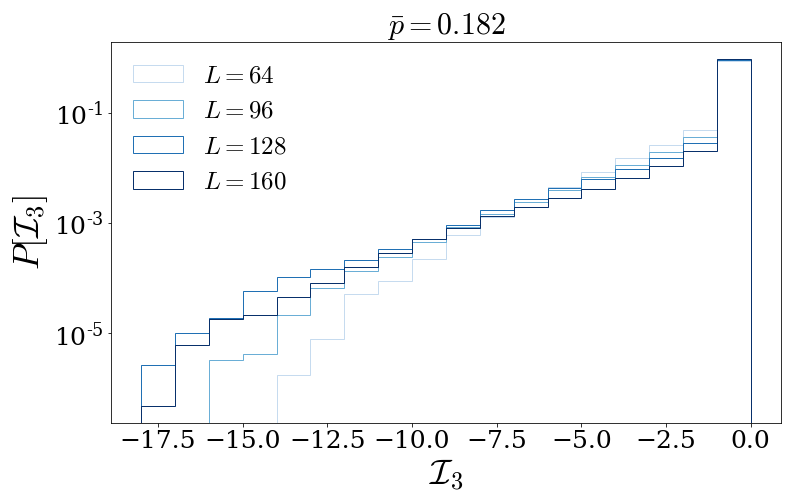}}
    \caption{\label{fig:I3_dist}
        Probability distribution of the tripartite mutual information,
        $P[\mathcal{I}_3]$, in the volume law and area law phases. Features to note are
        the broadening of the distribution and the weight at zero as a function
        of the system size.
    }
\end{figure}

In this section, we show how the distribution of the tripartite mutual information can be used as an estimate of the critical point.
In Fig.~\ref{fig:I3_dist}, we see that $P[\mathcal{I}_3]$ has
broad tails in the volume law phase which compress towards zero as one moves
into the area law phase. Looking at a particular $\avg{p}$ in the volume law
phase, we see that the distribution is broadening with increasing system size
and the weight of the distribution at zero, $P[\mathcal{I}_3 = 0]$, is decreasing. On the
other hand, in the area law phase, $P[\mathcal{I}_3 = 0]$ becomes $L$ independent and
approaches 1.
We propose the weight of the distrubtion at zero as a way to identify the
critical point since it does not require knowing the scaling of $\mathcal{I}_3$
only the assumption that it crosses at $\avg{p}_c$.

\subsection{Alternative identification of the critical point in the disordered model}
\begin{figure}
    \centering
    \subfloat[\label{fig:I3_La}]{\includegraphics[width=0.3\linewidth]{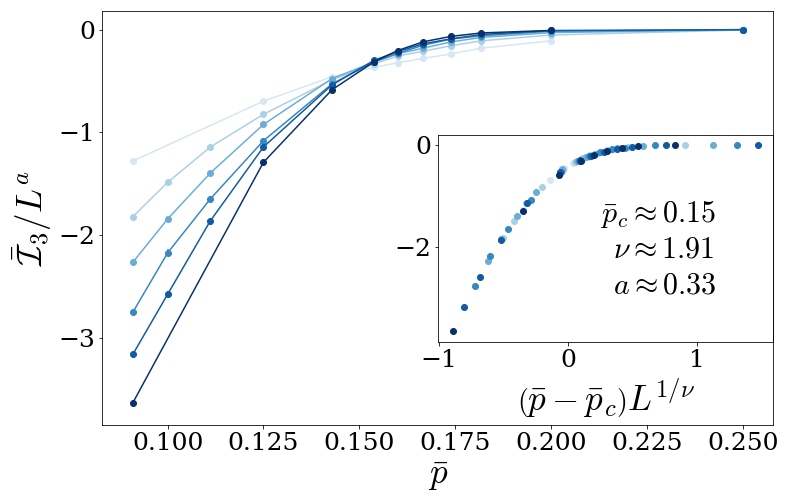}}
    \subfloat[\label{fig:I3_sqrtL}]{\includegraphics[width=0.3\linewidth]{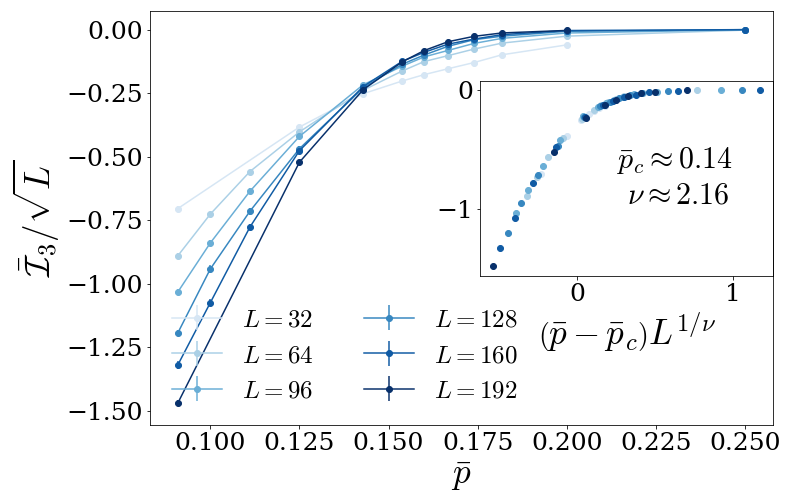}}
    \subfloat[\label{fig:I3_La_nu2}]{\includegraphics[width=0.3\linewidth]{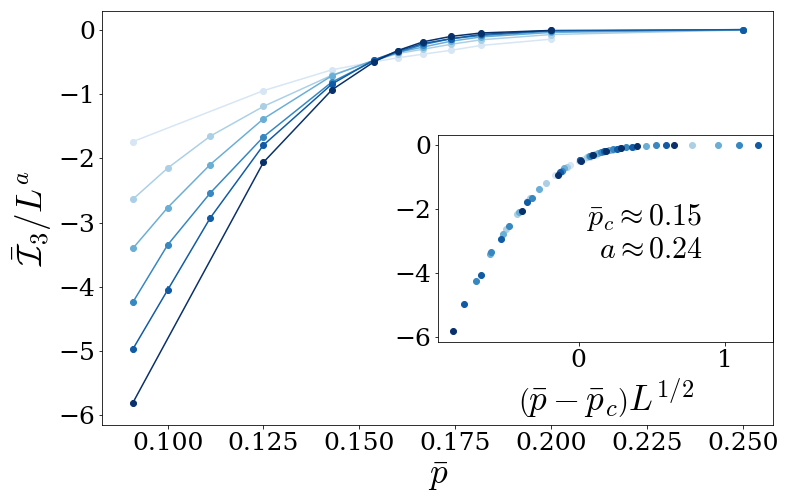}}
    \caption{
        Data collapse of the tripartite mutual information $\avg{\mathcal{I}}_3$. The free parameters
        $a$, $\avg{p}_c$, and $\nu$ are chosen as specified in the text and such that it minimizes the $\chi^2$.
    }
\end{figure}

In this section, we focus on identifying the transition using a data collapse of $\avg{\mathcal{I}}_3$ with the less constrained scaling ansatz,
\begin{equation}
    \avg{\mathcal{I}}_3 \sim L^a F[L^{1/\nu}(\avg{p}-\avg{p}_c)],
\end{equation}
to account for any possible $L$ dependence at the critical point.
We consider three cases of the scaling function:
\begin{itemize}
    \item In the most general case we choose $a, \avg{p}_c, \nu$ to be free parameters
        (Fig.~\ref{fig:I3_La})
    \item Based on the $\sqrt{L}$ scaling of $S(x,t)$ at the critical point we
        fix $a = \frac{1}{2}$ and choose $\avg{p}_c, \nu$ to be free parameters
        (Fig.~\ref{fig:I3_sqrtL})
    \item Based on the analytical results for large on-site Hilbert space
        dimension we fix $\nu = 2$ and choose $a, \avg{p}_c$ to be free parameters
        (Fig.~\ref{fig:I3_La_nu2})
\end{itemize}
The results for each of the methods are shown in Table~\ref{tbl:I3_pcs}.
In all instances we find similar values for $\avg{p}_c$, $\nu$, and $a$ that are consistent with the results quoted in the main text that were obtainted from the data collapse of the distribution.
\begin{center}
\begin{tabular}{ c c c c }
 \hline
 \hline
 & $\avg{p}_c$ & $\nu$ & $a$ \\
 \hline
    Case 1 & 0.15 & 1.91 & 0.33 \\
 \hline
    Case 2 & 0.14 & 2.16 & 0.5\\
 \hline
    Case 3 & 0.15 & 2 & 0.24 \\
 \hline
    $P[\mathcal{I}_3 = 0]$ & 0.14 & 2.21 & --\\
 \hline
 \hline
\end{tabular}
\label{tbl:I3_pcs}
\end{center}

\section{Percolation}
\begin{figure}
    \centering
    \subfloat[]{\includegraphics[width=0.32\linewidth]{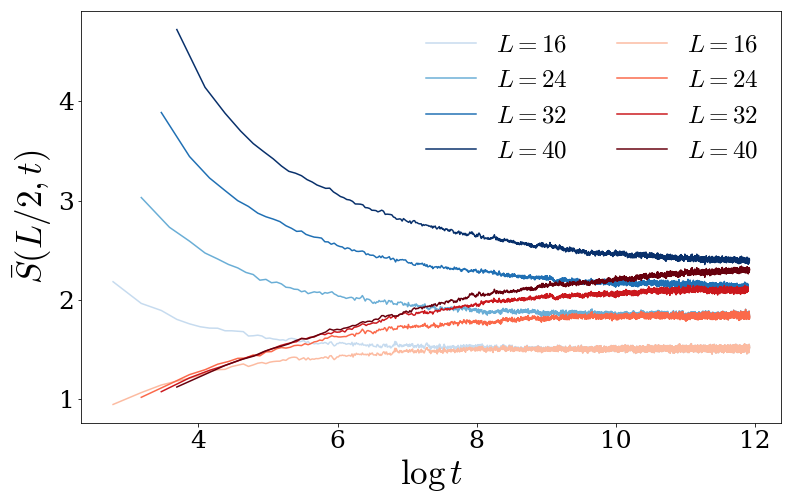}}
    \subfloat[]{\includegraphics[width=0.32\linewidth]{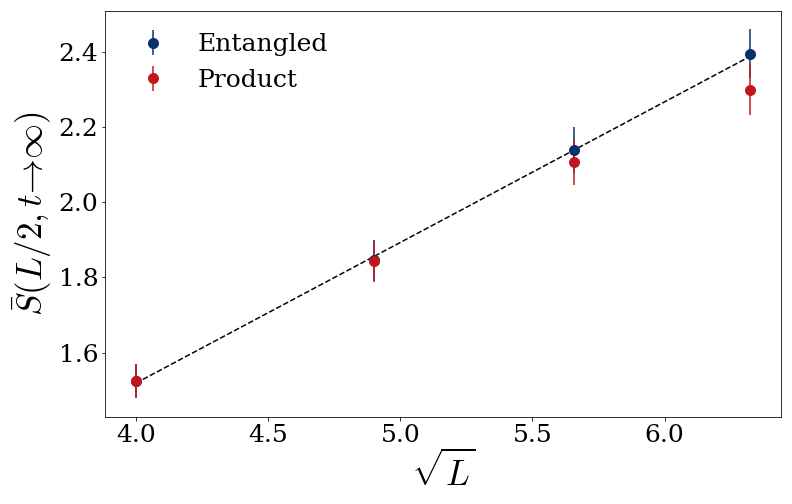}}
    \subfloat[]{\includegraphics[width=0.32\linewidth]{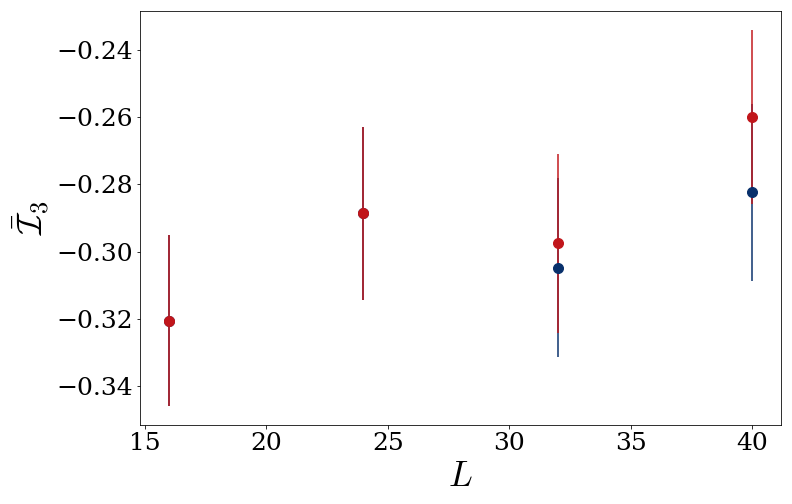}}
    \caption{
    (a) Half-cut entanglement entropy growth at the critical point of the percolation model for entangled (blue) and product (red) initial states.
    (b) At the critical point, the late time entanglement entropy behaves as $S\sim\sqrt{L}$ as shown by the black dotted line fit. This result matches the behavior of the entanglement entropy in the disordered circuit model.
    (c) In contrast with the disordered circuit model, the percolation $\avg{\mathcal{I}}_3$ is independent of the system size at the critical point.
    }\label{fig:percolation}
\end{figure}

The introduction of static disorder in a generically random model does not change the established connection of the Hartley entropy $S_0$ with bond percolation on a tilted square lattice \cite{skinner2019measurement}.
The connection to percolation allows us to perform classical calculations to characterize this transition as well.
Furthermore, bond percolation on a square lattice has a duality between cut and uncut bonds, allowing us to pin the transition at $\avg{p}_c = 1/2$ provided the distribution of probabilities is symmetric about $1/2$.
Therefore, we need to use a different distribution than what is used in the main paper. 
The distribution we use is
\begin{equation}
    P(x)dx = \frac{D + (D-1)(1-2x)^2}{(D - (D-1)(1-2x)^2)^2}dx, \label{eq:static_distribution_perco}
\end{equation}
where $D$ controls the degree of disorder. 
While this family of distributions naturally sit at the critical point, we can move away from the critical point by using a function $x(y) = (1-q)y/(q + (1-2q) y)$ and the distribution changes
\begin{equation}
    P(y) dy = P[x(y)] \frac{dx}{dy} dy =  \frac{D + (D-1)(1-2x(y))^2}{(D - (D-1)(1-2x(y))^2)^2} \frac{q(1-q)}{(q + (1-2q) y)^2}dy,
\end{equation}
where now $q$ tunes us away from symmetric distributions.

\subsection{Long cylinder percolation algorithm}
\begin{figure}
    \centering
    \includegraphics[width=\linewidth]{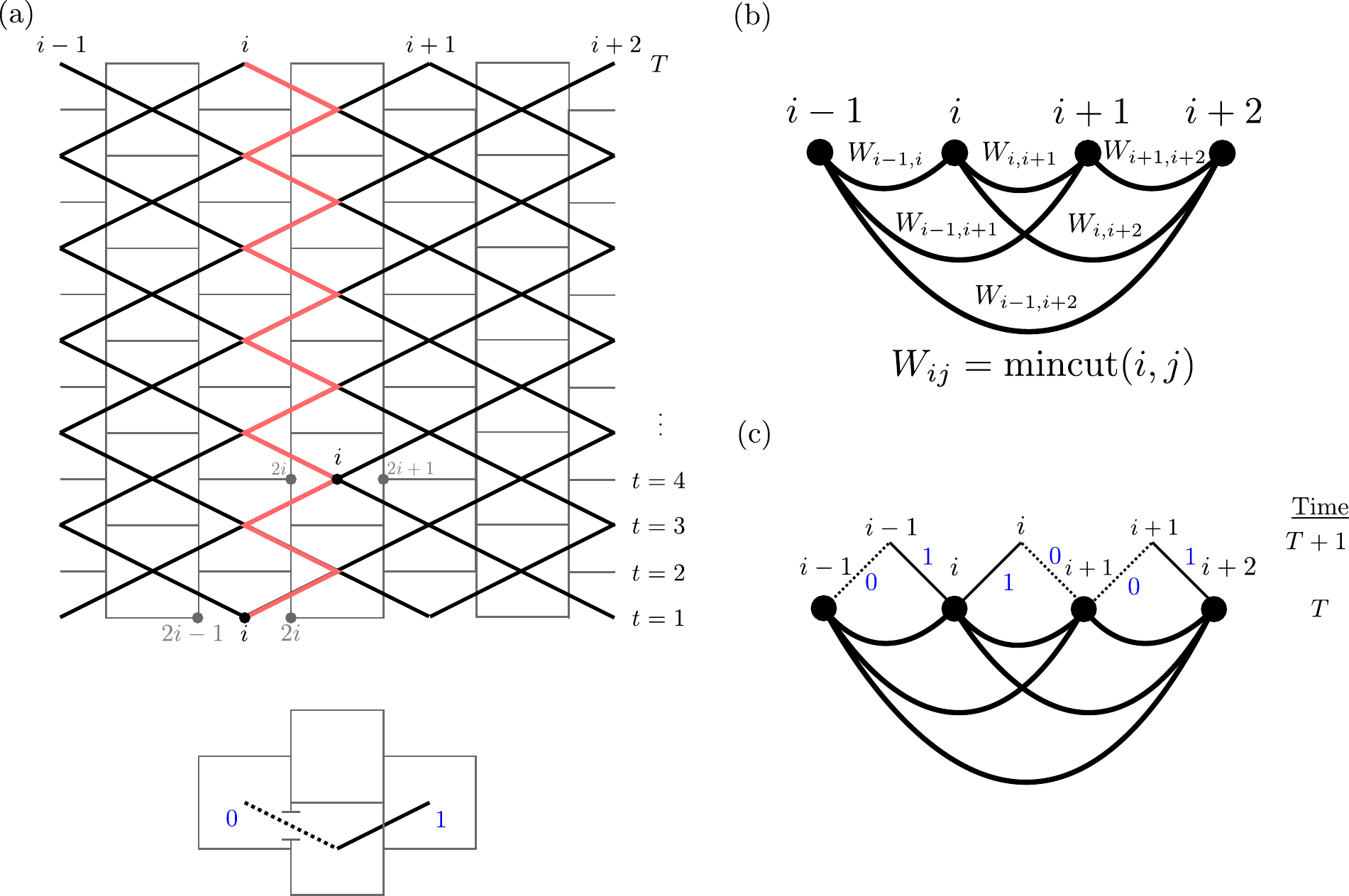}
    \caption{(a) The underlying qubit circuit (gray lines) with the overlaid tilted square lattice. Notice how vertical ``zig-zags'' represent the individual qubits now (red line is qubit $2i$). The bottom shows how we represent measurements (a dashed line on the dual tilted square lattice). (b) The leading edge graph where the vertices at time $T$ are fully connected but with weights that are equal to the shortest path on the tilted square lattice originally. (c) The time-step graph created from adding on the next layer along with bonds that are either weight 1 or 0. This new graph has shortest weighted paths that are the same length as the tilted square lattice. }
    \label{fig:perc_algo}
\end{figure}

In order to use percolation to very late ``times,'' we need to find the minimal cuts on very long cylinders at the edge of the system.
Standard percolation algorithms build the entire percolating network as a sparse matrix which implies $O(L^2T^2)$ operations; however, we can focus on the leading edge, calculating \emph{all} minimal cuts on that edge for $O(L^3 T)$ operations.

The basic idea of the algorithm is given in Fig.~\ref{fig:perc_algo}. After constructing the dual lattice, the minimal cut has been reduced to a shortest-path algorithm with some bonds that have weight $0$ (cut bonds) and some that have weight $1$ (uncut bonds). 
In this language, a cut starting between sites $2i-1$ and $2i$ (or $2i$ and $2i+1$) and ending between sites $2j-1$ and $2j$ ($2j$ and $2j+1$) during even (odd) time steps  is represented by a shortest path between $i$ and $j$ on the dual lattice [see Fig.~\ref{fig:perc_algo}(a)].
Once calculated, we replace our dual graph with a graph where each $i$ and $j$ on the leading edge is connected by a bond whose weight is precisely its shortest path.
If we then build up the next time step, we can use those shortest paths along with new bonds connecting the layers to compute new shortest paths, without needing the full dual lattice.

Some definitions: The \emph{dual graph} at time $T$ is defined by the vertices  labeled by $(i,t)$ for a position $1 \leq i \leq L/2$ and a time $1 \leq t \leq T$ as in Fig.~\ref{fig:perc_algo}(a).
Furthermore, it is a weighted graph where the weight is 0 if it crosses a broken bond (measurement) and 1 if it crosses an intact bond (no measurement).
We define $\mincut_{1\leq t \leq T}(i,j)$ as the length of the shortest path on the dual graph at time $T$ between $(i,T)$ and $(j,T)$.
Next, we define the \emph{leading-edge graph} at time $T$ as a complete, weighted graph on $L/2$ vertices with the weights defined by $W_{ij} = \mincut_{1\leq t \leq T}(i,j)$, see Fig.~\ref{fig:perc_algo}(b).
Last, we define the \emph{time-step graph} from $T$ to $T+1$ as the graph made by taking the leading-edge graph at time $T$ and connecting $L/2$ new vertices labeled $(i, T+1)$ with connectivity and weights inherited from the dual graph at time $T+1$, see Fig.~\ref{fig:perc_algo}(c).

The algorithm is then based on the following simple theorem:

{\bf Theorem} The quantity $\mincut_{1\leq t \leq T+1}(i,j)$ is equal to the length of shortest path of $(i,T+1)$ to $(j,T+1)$ on the time-step graph from $T$ to $T+1$.

\emph{Proof}: 
Let the shortest path on the dual graph of time $T+1$ from vertex $(i,T+1)$ to $(j,T+1)$ be labelled by consecutive bonds $(b_1,b_2, \cdots, b_N)$.
We can break this up into paths that exist purely on bonds that go from time $T$ to $T+1$ and bonds that go between all other times.
Breaking up those paths, the geometry of the tilted square lattice implies the path is equivalent to the ordered set $(\ell_1, r_1, \ell_2, r_2, \ldots, r_{M-1}, \ell_M)$ where $\sum_j |\ell_j| + \sum_k |r_k| = N$.

If $r_k$ goes from point $(a, T)$ to $(b, T)$, we first need to show $|r_k| = \mincut_{1\leq t \leq T}(a,b)$.
This is easily established by contradiction: if $|r_k| > \mincut_{1\leq t \leq T}(a,b)$, then we can just choose the minimal path since no vertices on the $T+1$ layer are within $r_k$, contradicting our statement that the original path $(b_1, \cdots b_N)$ was the shortest path. 
On the other hand, if $|r_k| <  \mincut_{1\leq t \leq T}(a,b)$, then since $r_k$ does not include vertices on the $T+1$ layer, it would be a path on the dual graph of time $T$ that is shorter than the $\mincut_{1\leq t \leq T}(a,b)$, contradicting its definition.

Finally, on the time-step graph from $T$ to $T+1$ take the path $(\ell_1, \tilde r_1, \ell_2, \cdots \tilde r_{M-1}, \ell_M)$ where $\tilde r_2$ goes directly from $(a,T)$ to $(b,T)$ on the bond of weight $\mincut_{1\leq t \leq T}(a,b)$. 
By definition of the weights on the time-step graph, this path has a length equal to the length on the dual graph at time $T+1$.
If a shorter path can be created, we can similarly break it up $(\ell_1', \tilde r_1', \ell_2', \cdots \tilde r_{M'-1}', \ell_{M'}')$ however, this corresponds to a path where $\tilde r_j'$ is replaced by a path on the dual graph of time $T$ $r_j'$ that is of the same length, and therefore we have constructed a path on the dual graph of time $T$ that is shorter than our original, a contradiction, proving the theorem. 

\vspace{6pt}

With this theorem, we need only keep track of $W_{ij}(T) = \mincut_{1\leq t\leq T}(i,j)$ and the connections
\begin{equation}
 C_{ij}(T+1) = \begin{cases}
   1 & \text{if $(i,T)$ to $(j,T+1)$ crosses an unmeasured bond,} \\
   0 & \text{if $(i,T)$ to $(j,T+1)$ crosses a measured bond,} \\
   \infty & \text{if not adjacent.}
 \end{cases}
 \end{equation}

The full weighted bond matrix for the time-step graph is then
\begin{equation}
  \mathcal W(T+1) = \begin{pmatrix} W(T) & C^T(T+1) \\
  C(T+1) & D
  \end{pmatrix},
\end{equation}
where $D_{ii} = 0$ and $D_{ij} = \infty$ if $i\neq j$.
Since $\mathcal W$ is $L \times L$, we can now perform the Floyd-Warshall shortest path algorithm to find all $W_{ij}(T+1)$. In this way, all minimal cuts can be constructed via an initial minimal cut matrix $W_0$
\begin{enumerate}
    \item $W(0) \leftarrow W_0$
    \item For $t=1$ to $T$: Generate $C(t)$ matrix, construct $\mathcal W(t)$ with $W(t-1)$ and $C(t)$, and perform Floyd-Warshall on $\mathcal W(t)$ to find $W(t)$.
    \item return $W(T)$.
\end{enumerate}
The expensive part of this algorithm is the Floyd-Warshall step which scales as $L^3$, making the run time for this algorithm $O(L^3 T)$. In particular, for exponential times, this algorithm wins over algorithms that scale like $T^2$.
This is why we call it the \emph{long-cylinder percolation algorithm}.

\subsection{Percolation results}

We focus on using the distribution in the strongly disordered $D=10$ case. 
The percolation probabilities apply to bonds that correspond to the qubits they correspond to; so for instance, all the red bonds in Fig.~\ref{fig:perc_algo}(a) have a probability $p_i$ drawn from the distribution Eq.~\eqref{eq:static_distribution_perco}.
Here, we can simulate a fully entangled state by initializing the weights $W_0$ as purely off-diagonal with each diagonal $W_{i, i + l} = 2l$ (with indices mod $L/2$), and a product state just uses $W_0 = 0$.

Defining the half-cut entropy as $S(L) = \mincut_{1\leq t \leq T}(i, i + L/4)$ (the Hartley entropy), we obtain Fig.~\ref{fig:percolation}(a) and see it takes exponentially long time to saturate the half-cut entropy.
Furthermore, by looking at Fig.~\ref{fig:percolation}(b), we can clearly see scaling that $S(L) \sim \sqrt{L}$ as is expected for the strong disorder.
Lastly, we can also compute $\mathcal I_3$ with a combination of minimal cuts and observe that it is roughly $L$ independent within error bars, see Fig.~\ref{fig:percolation}(c).

\section{Statistical mechanics model and RSRG}

Let us consider a monitored Haar quantum circuit with measurements occurring with a site-dependent probability $p_i$. Entanglement properties of quantum trajectories can be mapped onto a classical replicated statistical mechanics model whose degrees of freedom are permutations of the replicas~\cite{jian2020measurement,bao2020theory}. The quenched probabilities of measurement then  translates into some columnar quenched disorder in the statistical mechanics description.  We remark that the analysis in this section is not completely rigorous as we do not formally prove that our analytic continuation used in the replica limit is unique.  However, despite the lack of a rigorous justification, there is strong evidence from known results for clean Potts models that one can prove such a uniqueness result \cite{PaulMartinBook}.  Instead, we attribute a plausible explanation for the numerically observed differences between the percolation and Clifford critical theories mentioned in the main text to finite-$q$ corrections that could potentially affect the critical behavior as in the clean case~\cite{jian2020measurement}.

In the limit of infinite onsite Hilbert space dimension ($q \to \infty$), we have a Potts model with $k!$ states where $k$ is the number of replicas, defined on a tilted square lattice~\cite{jian2020measurement,bao2020theory}
\begin{equation} \label{eqPottsZ}
Z = \sum_{g_i \in S_k} \prod_{\langle i,j\rangle} \left( (1-p_{\langle i,j\rangle}) \delta_{g_{i},g_{j}} + p_{\langle i,j\rangle}\right).
\end{equation}
The tilted square lattice is a square lattice rotated by $45$ degrees. Each site of the tilted square lattice corresponds to a unitary gate in the original circuit. The degrees of freedom (``spins'') of the statistical model are permutations $g_i \in S_k$  defined on those sites, with fixed boundary conditions on the top layer of the circuit set by the entanglement properties one is interested in~\cite{jian2020measurement,bao2020theory}.
Here, the measurement probabilities $p_{\langle i,j\rangle}$ are inhomogeneous only in the spatial direction, and constant in the vertical (imaginary time) direction. The degrees of freedom in the Potts model are permutations, but this does not matter in that limit, beyond the fact that there are $k!$ states with $k \to 1$ in the replica limit.

The replica limit corresponds to percolation with columnar quenched disorder, which we here analyze using strong disorder RG techniques. It is known that the critical random Ising ($k=2$) and $k!$-state Potts model for $k!$ integer are described by infinite random fixed points~\cite{FisherRSRGIsing,SenthilMajundarPotts}. In order to take the replica limit $k \to 1$ in a controlled way, we derive an algebraic real space renormalization group (RSRG) approach to the Potts model, which allows us to analytically continue the number of states $k!$ to any real number.   

The first step is to notice that the transfer matrix of the Potts model~\eqref{eqPottsZ} on the tilted square lattice can be written in terms of operators $e_i$, which generate the so-called {\em Temperley-Lieb} (TL) algebra~\cite{PaulMartinBook}. This algebraic formulation will allow us to work with any representation of the Potts model, in terms of spins, clusters or loop gas, which in turn will allow us to take the replica limit in the end. The TL algebra consists of all the words written with the $N-1$ generators $e_i$ ($1 \leq i \leq N-1$), subject to the relations
\begin{subequations} \label{TLdef}
\begin{eqnarray}
\left[ e_i , e_j \right] &=&0 \ {\rm for} \ \left|i-j \right| \geq 2 ,\\
e_i ^2 &=& \sqrt{k!}  e_i ,\\
e_i e_{i \pm 1} e_i &=& e_i,
\end{eqnarray}
\end{subequations}
Up to a normalization factor $1/\sqrt{k!}$, the operators $e_i$ can be thought of as projectors. For example, for $k=2$ we have an Ising model and the TL generators read
\begin{subequations}\label{eqeisIsing}
\begin{eqnarray}
e_{2i-1} & = & \frac{1}{\sqrt{2}} \left( 1 + \sigma_i^x \right), \notag \\
e_{2i} & = & \frac{1}{\sqrt{2}} \left( 1 + \sigma_i^z \sigma_{i+1}^z \right).
\end{eqnarray}
\end{subequations}
For the $k!$-state Potts model, we have
\begin{subequations}\label{eqeisPotts}
\begin{eqnarray}
e_{2i-1} & = & \frac{1}{\sqrt{k!}} \left( 1 + \sum_{g_i \neq g_i^\prime}  \left| g_i \rangle  \langle g_i^\prime \right| \right), \notag \\
e_{2i} & = & \sqrt{k!} \ \delta_{g_i,g_{i+1}}.
\end{eqnarray}
\end{subequations}

Instead of working with a transfer matrix, it turns out to be more convenient to consider some anisotropic Hamiltonian limit. In the language of the original monitored circuits, this corresponds to considering the limit of continuous time (gates close to the identity), and weak measurements. 
Using the usual classical to quantum mapping, the universal properties of this statistical mechanics model can be inferred from the 1+1d effective Hamiltonian 

\begin{equation} \label{eqHTL}
H= - \sum_{i=1}^{N-1} J_i \frac{e_i}{\sqrt{k!}},
\end{equation}
where $J_i >0$ are random positive parameters, that are related to the original random probabilities $p_i$ in a limit of continuous time and weak measurements. Note that eq.~\eqref{eqHTL} combined with~\eqref{eqeisIsing} and~\eqref{eqeisPotts} corresponds to the familiar Hamiltonians for the Ising and Potts quantum chains. 
Criticality is obtained by enforcing  the same distribution of the couplings $J_i$ on odd and even bonds, which in terms of the original probabilities can be achieved through a statistical symmetry $p_i \leftrightarrow 1 - p_i$ as in the percolation problem above. 
Note that at this stage, the Potts model is formulated purely algebraically, and the number of states $k!$ can be analytically continued to a real number. In particular, we can consider a representation of the Potts model in terms of Fortuin-Kasteleyn clusters~\cite{PaulMartinBook}, where each cluster carries a Boltzmann weight $k!$: this corresponds to a different representation of the Temperley-Lieb algebra, where $k$ can be tuned continuously. The algebraic RSRG approach we derive below applies to any representation. 

As we will now show, the groundstate properties of the system can be understood in terms of a RSRG approach that can be carried out using only the commutation relations~\eqref{TLdef} of the TL algebra. As usual within the RSRG approach, we identify the strongest bond $\Omega = J_i$ of the chain, solve the corresponding local Hamiltonian $H_0 = - \Omega e_i/\sqrt{k!}$ and deal with the rest of the Hamiltonian $V = H-H_0 =  - J_{i-1} e_{i-1}/\sqrt{k!}- J_{i+1} e_{i+1}/\sqrt{k!} + \dots$ perturbatively. The groundstate manifold is defined by the projector $P_i = e_{i}/\sqrt{k!}$ and we also define $\bar{P}_i=1-P_i$. The effective Hamiltonian in the groundstate manifold can be obtained using a Schrieffer-Wolff transformation $H_{\rm eff} = \mathrm{e}^{iS} H \mathrm{e}^{-iS}-H_0$ where $S$ is obtained perturbatively in $V$ by requiring $[P_{i} ,  \mathrm{e}^{iS}]=0 $, that is, by requiring that $U= \mathrm{e}^{iS}$ decouples the low and high energy sectors $P_i {\cal H} $ and $\bar{P}_i {\cal H} $ of the Hilbert space ${\cal H} $. One then finds $S=\frac{1}{i \Omega} (\bar{P}_i V P_i - P_i V \bar{P}_i) + \dots$ and the resulting effective Hamiltonian is given by $H_{\rm eff}=P_i (V + [i S,V] - \frac{1}{2} \lbrace S^2,H_0 \rbrace + S H_0 S + \dots)P_i$. Using~\eqref{TLdef}, we find the first order term $P_i V P_i = -(J_{i-1}+J_{i+1})/k! P_i$ which acts as a constant in the subspace generated by $P_i$. The second order term then reads $P_i ([i S,V] - \frac{1}{2} \lbrace S^2,H_0 \rbrace + S H_0 S + \dots)P_i = -1/\Omega \times P_i V \bar{P}_i V P_i$ which yields, using~\eqref{TLdef}
\begin{equation}
H^{P_i}_{\rm eff} = C P_i - \frac{2 J_{i-1} J_{i+1}}{k! \Omega} \frac{\tilde{e}_{\rm eff}}{\sqrt{k!}},
\end{equation}
where $C$ is a constant that will renormalize the energy: the energy shift after the decimation is $-\Omega+C$, with $C=-\frac{J_{i+1}+J_{i-1}}{k!}+\frac{1}{\Omega}\left(\frac{(J_{i+1}+J_{i-1})^2}{(k!)^2}-\frac{J_{i+1}^2+J_{i-1}^2}{k!} \right)$, and 
\begin{equation}
\tilde{e}_{\rm eff} = \frac{1}{\sqrt{k!}} e_i e_{i-1} e_{i+1} e_i,
\end{equation}
is an effective TL generator is the low-energy subspace $P_i {\cal H} $. It is indeed straightforward to verify that $\tilde{e}_{\rm eff}^2=\sqrt{k!} \tilde{e}_{\rm eff}$, $ \tilde{e}_{\rm eff} e_{i\pm 2} \tilde{e}_{\rm eff}- \tilde{e}_{\rm eff}=0$ and $P_i ( e_{i\pm 2}\tilde{e}_{\rm eff} e_{i\pm 2}- e_{i\pm 2}) P_i=0$. This simple  calculation allows us to derive the expression of the effective coupling
\begin{equation} \label{eqJeff}
J_{\rm eff} = \frac{2}{k! }\frac{J_{i-1} J_{i+1}}{\Omega},
\end{equation}
purely algebraically. This process can be iterated by identifying the next largest coupling in the chain, and decimating it in the same way. The recursion relation~\eqref{eqJeff} agrees with previous results in the case where $k!$ is an integer~\cite{FisherRSRGIsing,SenthilMajundarPotts}, but the upshot of the above algebraic approach is that it allows us to analytically continue to $k$ real. In particular, this recursion relation also applies in the replica limit $k \to 1$, where it describes percolation in media with columnar disorder.

For any value of $k!$ and for strong enough initial randomness in the couplings $J_i>0$, the recursion relation~\eqref{eqJeff} is known to flow to an infinite randomness fixed point with space time scaling $\log t \sim \sqrt{\ell}$. The scaling properties discussed in the main text then follow from standard RSRG results~\cite{FisherRSRGIsing,FisherRSRGXXZ,IGLOI2005277,REFAEL2013725}. Those predictions also agree with earlier results on percolation in media with columnar disorder~\cite{PercolationRandom}.

\end{document}